**Sight, smell and more: What cues do free-ranging dogs use for decision-making while scavenging?**

Rohan Sarkar[1], Tuhin Subhra Pal[1], Sharmistha Maji[1], Srijaya Nandi[1], Aesha Lahiri[1], Ashim Kumar Basumatary[1], Arpitha GP[2,3], Nakul Wehare[4,5], Nudrat Jahan[6,7], Sharayu Jakhete[8,9], Srinwanti Bandyopadhyay[10], Rajdip Adhikary[11,12], Tithi Paul[13,10], Shreya Majumdar[14], Anindita Bhadra[1*]

[1] Department of Biological Sciences, Behaviour and Ecology Lab, Indian Institute of Science Education and Research Kolkata, Kolkata, West Bengal, India

[2] Department of Biology, Hill Lab, Auburn University, Alabama, USA

[3] Department of Applied Zoology, Mangalore University, Mangalore, Karnataka

[4] Department of Ecology and Evolutionary Biology, Princeton University, Princeton, New Jersey, USA

[5] Indian Institute of Science Education and Research (IISER)- Pune, Pune, Maharashtra, India

[6] Department of Neuroscience, National Brain Research Centre, Manesar, India

[7] Regional Institute of Education, Bhubaneswar, Odisha, India

[8] Huon Aquaculture, Hobart, Tasmania, Australia

[9] Ecological Restoration and Reef Adaptation Programme, Australian Institute of Marine Science, Queensland, Australia

[10] Department of Zoology, Visva Bharati University, Shantiniketan, India

[11] Hester Biosciences Limited, Ahmedabad, Gujarat

[12] Department of Biochemistry and Biophysics, University of Kalyani, Kalyani, India

[13] Department of Ecotoxicology, Institute for Environmental Sciences (iES), Rheinland-Pfälzische Technische Universität Kaiserslautern-Landau, Landau, Germany

[14] Department of Cognitive Neuroscience, Indian Institute of Science Education and Research, Berhampur, Odisha, India

**Abstract**

Finding food is a fundamental activity for survival of all living organisms. Free-ranging dogs have been known to use their olfaction to assess the quality and type of available food but their use of visual ability in foraging is not well-documented. In the current study, we seek to remedy that by testing free-ranging dogs in a food-based choice test. We tested whether the dogs implemented hierarchical or synergistic usage of cues while finding food. We found limited prioritization of olfactory cues over visual cues in dichromatic choice tests but in phases with similar perceptual elements, the sensory choice was not clear. Furthermore, free-ranging dogs display a dynamic decision-making in unpredictable urban environments adopting a "good-enough" strategy during foraging. They prefer speed over accuracy, settling for intermediate quality food if their preferred food item is not available. These dogs also displayed left-bias during food choice. In multi-sensorial, natural setting multiple modulators like environmental noise, risk, and internal perceptual elements apart from food cues seem to be affecting the decision-making in dogs.

1. **Introduction**

In complex and ever-changing environment, animals use both environmental and sensory information to forage for food. Foraging efficiency depends on detecting, discerning, and reacting to predictive signals from food. Animals use different cues to locate food, such as visual, olfactory, tactile, and spatial cues [1], [2], [3], [4], [5] in a variety of ways. When using environmental signals for foraging decision-making, animals may use one particular cue, may display sensorial hierarchy, or use multiple modalities in tandem to locate food [6], [7]. Sometimes, multiple cues may even provide conflicting information [8]. The ability to use their sensory modalities flexibly is an even more important adaptation for animals in urban environments where the selection of relevant information comes under interference from sensory pollutants like human generated light, noise, and volatile chemicals [9]. Understanding the foraging behaviour of an animal requires knowledge about their perceptual, sensory, and cognitive abilities, the environment they inhabit, and how they detect, assess, and use the information available to them.

Broadly, foraging consists of two stages- exploration and consumption of food. Exploration consists of searching and detecting food. This consists of multiple decisions like assessing food quality, discriminating between available food, and balancing preferences. Among the multiple signals available to animals, sight and scent are two of the most important signals used in food detection and selection. Studies on the role of visual and olfactory cues in foraging has shown the dual importance of these cues by themselves or in combination [10], [11], [12], [13]. The relative importance of visual and olfactory signal usage in food detection also depends on the perceptual abilities and the properties of the food they frequently eat. For example, vervet monkeys possess trichromatic vision and primarily rely on visual signals to find food whereas the Indian fulvous fruit bat depends on olfactory signals to find fruits [11], [14]. Visual and olfactory cues can be used to assess the palatability and nutritional qualities of food at relatively longer distances [15], [16], [17], [18]. In terms of decision-making, these cues are energy efficient. Therefore, they play an important role in diet selection and foraging efficiency [19], [20]. Additionally, context and condition may dictate the prioritization of one sense over another. For example, visual cues are more effective at targeting large prey items during day and olfactory cues are preferred at night when visual cues are obscured [17], [21]. The habitat and foraging grounds of an animal also influences the usage of cues [13], [22]. Thus, multiple factors, both internal and external, influence the sensory prioritization or integration in animals during foraging.

Multi-sensory animals can use olfaction and vision synergistically to optimize their foraging decisions. Honey buzzards use a multi-modal foraging strategy, integrating vision and olfaction in identifying nutritious food, with the olfactory signal chosen over vision when the two senses provide conflicting information [23]. While the detection and consumption of high nutrient food using sensory systems point to optimized foraging decisions, failure to do so is not necessarily maladaptive. In ecologically-realistic scenarios, multiple environmental as well as internal, non-food signals are mixed with the food signals that could lead to context-dependent outcomes that don't always lead to selection of high nutrient food. In urban habitats, due to patchy availability of natural resources and high competition for food, urban animals tend to maximise their food intake whenever they can. This leads to more willingness and higher exploitation of energy rich, but nutrient poor anthropogenic food by urban animals as opposed to their wild counterparts[24], [25]. Thus, it is important to test sensory processing and food choices in realistic, multi-sensory settings to understand their function and outcomes in foraging decisions.

Another factor influencing foraging decisions is the lateralized processing of sensory information that leads to the preferential use of the sense organs on the left or right side of the body [26], [27], [37], [38]. Behavioural and their associative hemispherical lateralization confer certain fitness benefits. Animals with a high degree of visual lateralization are more efficient in food discrimination and predator detection [28], [29]. Compartmentalizing type-specific signal processing to specific sides of brains leads to more rapid and efficient decision-making [30]. Lateralization is also known to play a role in decisions related to food preference tests. In a study to assess palatability in relation to lateralization, horses were found to choose the food item on the right side, independent of the palatability of given feed [31].

Among domesticated animals, dogs are known for their olfactory acuity [32]. This is reflected by the large number of olfactory neurons present in their olfactory epithelium, between 220 million-2 billion [33], [34]. Historically, it is theorized that dogs were domesticated for their use in hunting where their superior sense of smell was an asset [35], [36]. In the present day, dogs are used for a variety of scent-detection tasks [37], [38], [39]. Despite their keen olfactory sense, the processing of such signals has certain limitations. Olfactory cues are limited by wind directions and the slow nature of diffusion [40]. There are also contradicting evidence to the olfactory capabilities of dogs such as when even well-trained search dogs failed to find their owners sitting nearby using only olfactory cues [41]. Comparatively, the visual system and perception in dogs is understudied. Dogs are known to be visual generalists with better functional adaptation in dim light [42], [43]. The visual acuity of a dog is inferior to that of a human [43]. Dogs are known to have dichromatic vision, similar to that of humans with red-green colour blindness [43], [44]. Dogs are able to visually discriminate large quantitative differences but perform poorly when the differences are small [45], [46]. Like other canids, dogs have also been found to prefer vision over olfaction when searching for hidden food [47], [48]. Studies on lateralization in dogs have found that they show a left gaze bias towards human faces, the right side of their brain is more responsive to threatening or alarming stimuli, right nostril bias to sniff at alarming odours, and they wag their tail to the right side for stimuli eliciting approach responses [49], [50], [51], [52]. Moreover, dogs show a population-level right-eye/left hemisphere bias when inspecting food [53].

The Indian free-ranging dog (FRD) presents an ideal model system to study the relative sensory reliance of vision versus olfaction in foraging decision-making in realistic multi-sensory ecological settings. These dogs live on the streets free of human supervision and have been a ubiquitous part of human settlements for millennia [54], [55], [56]. They subsist on a carbohydrate-rich diet but maintain a preference for meat in both controlled and natural conditions using a Rule of Thumb, "If it smells like meat, eat it first", applied through the Sniff-&-Snatch strategy (SNS) [57], [58]. These dogs are known to employ a random search optimization process during foraging and eventually transitioning into systematic foraging as more information became available [59]. The dogs assessed the quality and quantity of available food in these experiments through the integration of both visual and olfactory information, although olfactory sensing was dominant. They forage in a competitive environment where they have to compete not only with group members and conspecifics, but also other scavenger species [59], [60]. As a result, they are quick to respond to the presence of food and prioritize eating from less-than-optimal patches, if they happen to encounter it first [59]. They are also able to perform quantitative discrimination in a context-dependent, olfactory- dominated, visually supplemented manner [61]. Alternatively, in a novelty task, they depended more on visual cues than olfactory cues [62]. They have also shown preference towards yellow colour in a choice test and the preference for approaching yellow items supersedes their attraction for food, although the cause behind such non-optimal behaviour in the context of foraging decisions is not yet clear [63].

In this study, we aimed to understand how FRDs perceive multimodal stimuli in a natural, urban, unpredictable setting by estimating the relative roles of vision and olfaction in dog foraging behaviour. We carried out food based, choice-tests with FRDs to assess if untrained dogs prefer one modality over another or perceive them in combination while searching for food. We presented them with different sets of choices and the dogs' selection of a choice-item would help us understand the decision-making points behind the choices they made. We hypothesized that:

Hypothesis 1: In all cases, dogs were expected to approach and sniff the choices randomly.

Hypothesis 2: In unimodal or near-unimodal conditions, dogs were expected to eat the higher olfactory cue item more than the higher visual cue item. In multi-modal conditions, they were expected to eat the item that cumulatively provide the highest signal. This would translate to chicken being eaten more in both cases.

Hypothesis 3: Dogs would show quick, differential decision-making on the basis of age.

Hypothesis 4: There is no preference for a particular colour.

Since there has been no study to understand behavioural asymmetries in free-ranging dogs and the influence of such responses to preference tests, we also raised (and subsequently tested) the question of whether laterality influenced foraging decisions.

## 2. Methods

### 2.1. Study areas

The study was conducted across 72 field sites in 4 states of India (West Bengal, Maharashtra, Orissa, Karnataka). The field sites have been highlighted in the map (Figure 1). The study was carried out from 2018-2024.

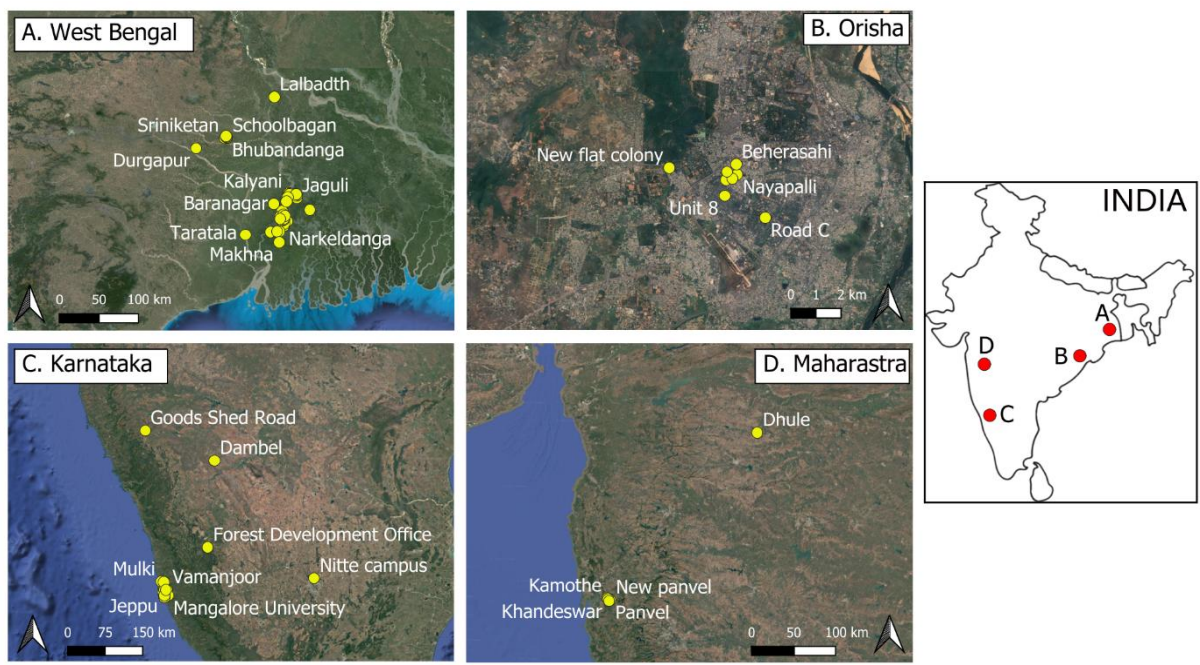

**Figure 1:** Map marking the states and towns where field sites were located

### 2.2. Subjects

We tested 1153 dogs in total, out of which 961 dogs (390 adults, 278 juveniles, 293 pups; 413 males, 541 females, 9 unknown) spread across the phases were included in the analysis. Dogs that did not successfully complete the familiarisation were not included (see Experimental Procedure). While the exact age of the dogs was not known, the age class could be determined through morphological characteristics [64]. Dogs were located randomly in different areas for the experiment. The experiment was carried out in different locations each day, or with a gap of at least a week for a location, to avoid resampling.

### 2.3. Experimental Procedure

Experimenters E1 and E2 would select a location and start walking. On encountering a dog, they would first conduct a familiarisation experiment. This consisted of showing a red box with a piece of biscuit (half of a Parle-G biscuits) inside to the dog. The box was placed in front of the dog (1-2m) with the lid loosely placed on top. This was to familiarise the dog with the set-up and reward system. The dog was given 1 minute to interact with the box and eat the biscuit. Only dogs that successfully completed this stage were allowed to carry out the trials. Each dog was subjected to three trials. In each of the trials, a two-way choice test was provided. This consisted of E1 placing two boxes (the box type being decided by the phase that was being conducted), 1m apart in front of the dogs (1-2m from the dog). The placement of the boxes was randomized. The boxes had their lids loosely placed on top. This had the effect of removing the visual cue in the black box but was also easy for the dogs to access the food once located. The dog was allowed a total of 1 minute to interact with the set-up. One of the boxes contained a chicken piece and the other contained one-half of a biscuit (a different biscuit than the one presented during the familiarisation to avoid association, Marie© biscuit). The experiment was stopped as soon as the dog ate the food item from one of the boxes. The dog was allowed to sniff both boxes if it had not eaten from any of the box before doing so.

### 2.4. Phases

There was a total of five phases - four for food-reward tests, and a control phase with no food. The phases were based on the cue that they provided the dogs. Each phase consisted of a specific type of choice test being administered to the dogs. The choice tests consisted of two boxes with two different food items kept in them. The four test phases and the respective signals from the boxes were as follows:

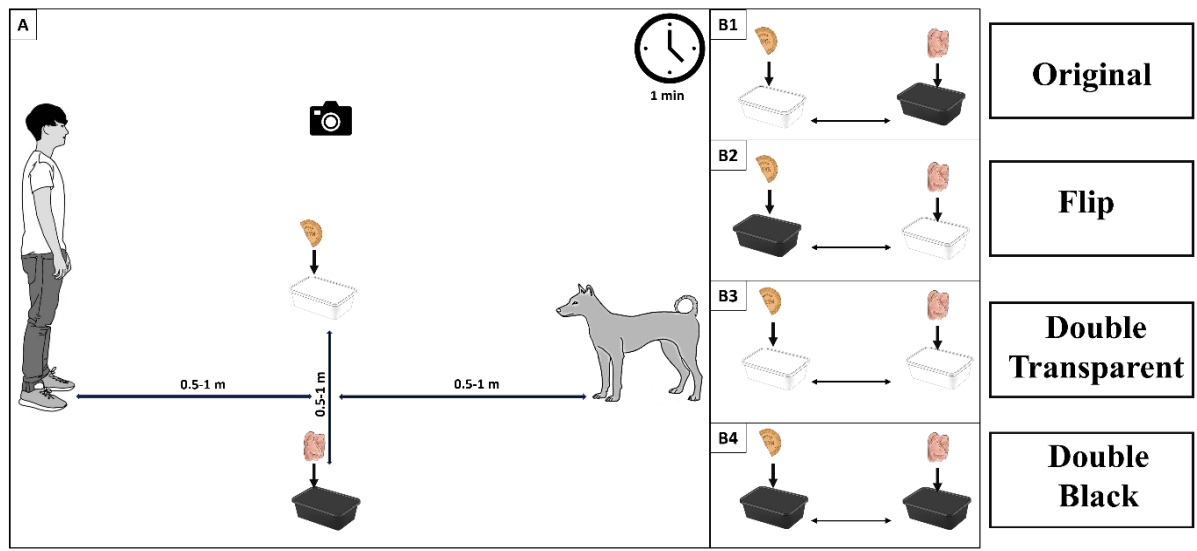

**Figure 2:** A. The experimental set-up and B (1 - 4). The combinations of boxes used for the different phases.

|  | Visual cue | Olfactory cue |
|---|---|---|
| Chicken in black box (Original phase) | ✗ | ✓ ✓ |
| Biscuit in transparent box (Original phase) | ✓ | ✓ |
| Chicken in transparent box (Flip phase) | ✓ | ✓ ✓ |
| Biscuit in black box (Flip phase) | ✗ | ✓ |
| Chicken in black box (Double black phase) | ✗ | ✓ ✓ |
| Biscuit in black box (Double black phase) | ✗ | ✓ |
| Chicken in transparent box (Double transparent phase) | ✓ | ✓ ✓ |
| Biscuit in transparent box (Double transparent phase) | ✓ | ✓ |

**Table 1:** The cues available in the four phases

The control phase with no food consisted of an empty black box and a transparent box to check for the effect of the colour of the boxes on the dogs themselves, if any.

### 2.5. Behaviours and Analysis

We considered two behaviours: sniffing and eating. A dog was said to be sniffing a box if it approached within 2-3cm from the edge of the box with the snout extended. Eating was defined as removing the lid and consuming the food inside.

Each phase was conducted on a different set of dogs. Thus, our study had a between-subjects design. All the analyses were done in R 4.3.1 [65]. We carried out chi-square goodness-of fit tests to compare frequencies against an expected ratio of chance. We carried out generalised linear mixed effects logistic regression for binomial data distribution using the lme4 packages [66] for between group comparisons. Multinomial mixed effects logistic regression [67] were carried out for multinomial data. Multiple models with different predictors and combinations thereof were run but we only reported predictors which had a significant effect on the outcome variable. We reported bias adjusted, model-estimated marginal means using the emmeans package [68], [69] and log odds estimates. Pairwise and multiple comparisons p-values were adjusted using the "false discovery rate" method and estimates were reported in log-odds scale, unless mentioned otherwise [70], [71]. Random effect variables were included in the models, despite some models having low variance components to maintain fidelity between our models and the data-generating process. We carried out time-to-event analysis through mixed effects cox regression using the coxme package and compared latencies through Kaplan—Meier survival (event) estimates [72]. 21% of the data was re-coded by a second rater naïve to the experimental hypothesis. The inter-rater reliability was measured through the Cohen's kappa coefficient for the categorical variables and were found to be 0.834 – 0.992. For the quantitative data, we used intraclass correlation estimates based on a single rating, absolute agreement, two-way mixed effects model and the scores was 0.987.

### 3. Results

#### 3.1. Do dogs initiate their sampling of the two boxes randomly or preferentially?

We compared the number of times the box containing the chicken was approached and sniffed first versus the biscuit containing box across the four phases. We only considered the first trial as dogs were naïve to the set-up and thus learning had no effect on their decision-making. The observed frequencies were compared against a ratio of 1:1. Chi square goodness of fit tests showed that dogs did no preferential sampling between the two food-containing boxes in any of the phases ($p > 0.05$).

### 3.2. Do dogs sample some boxes more than the others?

We investigated the first sampling event of dogs across three trials to check the food-box that was preferentially sniffed, if at all, within each phase and age group. The response variable, "first sniff choice" was binary (chicken/biscuit) referring to one of two boxes, identified by the food in them, that were first approached and sniffed. We ran 12 intercept-only hierarchical logistic regressions with random effects of "trial" nested within individual "dog id" nested within "bigplace" subsetting by phase and subsequently age (flipbox adult, flipbox juvenile, original pup and so forth)

firstsniffchoice ~ 1 + (1 | bigplace/dogid/trial)

The results, post p-value adjustment, showed that juveniles in flip box phase sniffed the chicken box (transparent box) less than the biscuit box (black box)(-0.44, p-value < 0.05) whereas juveniles in the original phase sniffed the chicken box (black box) more (0.45, p-value < 0.05).

### 3.3. Do dogs eat from some boxes more than the others?

We investigated the eating event of dogs across three trials to check if a particular box was preferentially eaten from, if at all, within each phase and age group. The response variable, "eatchoice" was binary (chicken/biscuit) referring to one of two boxes, identified by the food in them, that was eaten from. We ran 12 intercept-only hierarchical logistic regression with random effects of "trial" nested within individual "dog id" nested within "bigplace" subsetting by phase and subsequently age (flipbox adult, flipbox juvenile, original pup and so forth)

eatchoice ~ 1 + (1 | bigplace/dogid/trial)

The results, post p-value adjustment, showed that adults, juveniles, and pups in the original phase ate from the chicken box (black box) more than the biscuit box (transparent box) (Estimates: 0.79, 1.04, 0.55, respectively. All p-values < 0.05). Similarly, adults in the flip box phase ate from the chicken box (transparent box) more too (0.90, p-value < 0.05). No such preference was observed in double black and double transparent phase.

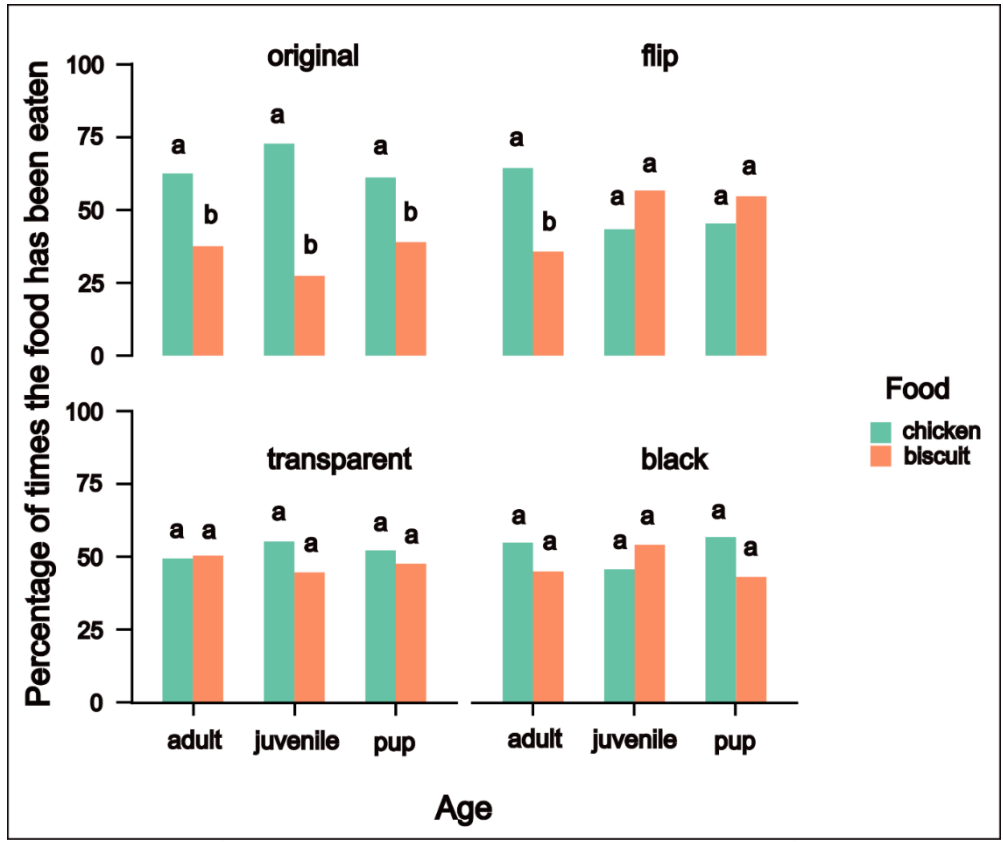

**Figure 3:** A bar graph showing the percentage of times a particular food has been chosen across age and phases

### 3.4. Do dogs show preference in their choice when they sniff both boxes before eating?

Out of 2334 instances of being provided with a choice across four phases, dogs sniffed both boxes 315 times before eating. Out of the 315, original, flip, double black, and double transparent phases had 108, 71, 73, and 49 instances of dogs sniffing both boxes respectively. We investigated whether doing so led to eating one type of food over the other. The response variable, "eatchoice" was binary (chicken/biscuit) referring to one of two boxes, identified by the food in them, that were eaten from. The predictor, "sniffedbothboxes" was binary (yes/no), referring to whether both the boxes were sniffed before initiating eating. We ran a hierarchical logistic regression with random effects of "trial" nested within individual "dogid" nested within "bigplace" (4A). Following this, we ran an intercept-only model after subsetting the data frame to only include the dogs who sniffed both boxes and then checked if they preferred chicken box more in all phases (4B).

Eatchoice ~ 1 + sniffedbothboxes + (1 | bigplace/dogid/trial)     … 4A

Eatchoice ~ 1 + (1 | dogid)                                       …4B

The model scored 0.85 on class separability. The regression results showed that sniffing both boxes resulted in higher consumption from chicken containing boxes (0.64, p-value < 0.0001). The results for the intercept-only model, post p-value adjustment revealed that chicken box was preferred more in original (3.17, p-value < 0.05) and flip (2.86, p-value < 0.05) phases whereas there was no preference in double black and double transparent phases.

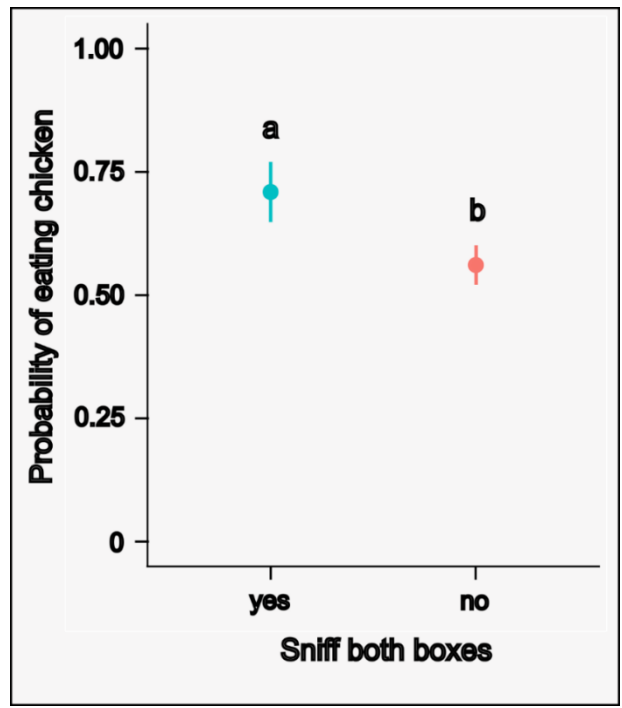

**Figure 4:** Model-estimated marginal means of the probability of eating chicken by a dog when they have sniffed both boxes versus not. The whiskers represent the 95% asymptotic confidence limits. The letters denote the presence of difference across the predictor, "sniffedbothboxes".

### 3.5. Do dogs follow the Sniff-&-Snatch strategy (SNS) for the chicken-containing box?

We tested if dogs were likely to follow SNS strategy (ref) for chicken (their preferred food) when they sniffed both boxes (and thus acquired complete information about the set-up). Only the dogs that sniffed both boxes were selected for this analysis. There were 315 instances of dogs having done so. The dogs could follow one of four foraging strategies: a) sniff chicken, sniff biscuit, eat biscuit (scsbeb), b) sniff biscuit, sniff chicken, eat chicken (snsyes), c) sniff biscuit, sniff chicken, eat nothing (sbscen), d) sniff chicken, sniff biscuit, eat nothing (scsben). Thus, the probability of snsyes being followed was one in four, 0.25. We ran an intercept-only multinomial regression with a random effect of "dogid". The simplicity of the model prevented us from adding nested grouping variables as that led to the inability of the model to converge. The response variable was "sns", that denoted one of the strategies being followed.

sns ~ 1 + (1 | dogid)

The results of the intercept only model showed us that dogs that sniffed both boxes ended up following the snsyes strategy more than the others: a) sbscen vs yes (-3.13, p-value < 0.05), b) scsbeb vs yes (-0.75, p-value < 0.05), c) scsben vs yes (-4.22, p-value < 0.05).

### 3.6. Does the side of a food/box placement determine its likelihood of being eaten?

Dogs approached the left side on 1404 instances and right on 930 instances when first exploring the set-up. Similarly, they ate from the left box on 1329 instances as compared to 979 instances from right. We tested two things – (a) whether dogs were likely to favour one side over other and (b) whether placing a food-containing box on a particular side made it more likely to be eaten. We ran an intercept-only model for the first case and a logistic regression for the second. Both models were hierarchical with random effects of "trial" nested within individual "dogid" nested within "bigplace". The intercept-only model had "eatchoiceside" as its response variable (binary: left/right) that denoted the side the dog approached and ate from. The binomial model had "eatfood" (food being replaced by chicken or biscuit depending on testing parameter) as the response variable (binary: yes/no) that denoted whether the food provided was eaten or not. The predictor variable, "positionfood" (binary: left/right; food being replaced by chicken or biscuit depending on the model) denoted the side, from the perspective of the dog, the food-containing box was placed in.

eatchoiceside ~ 1 + (1 | bigplace/dogid/trial) . . . Intercept-only model

eatfood ~ 1 + positionfood +(1 | bigplace/dogid/trial) . . . Binomial model

The binomial models had a class separability of 0.85. The results of the intercept only model showed us that the odds ratio of dogs approaching and eating from the left side were 1.34 times higher than what would be expected by chance (0.30, p-value = 0.0002). Furthermore, phase and age had no effect on the direction the dog chose to move and eat from. The logistic regression showed that placing either of the food-containing box on the left side made it more likely to be eaten for both chicken and biscuit ((+/-)0.72, p-value < 0.001).

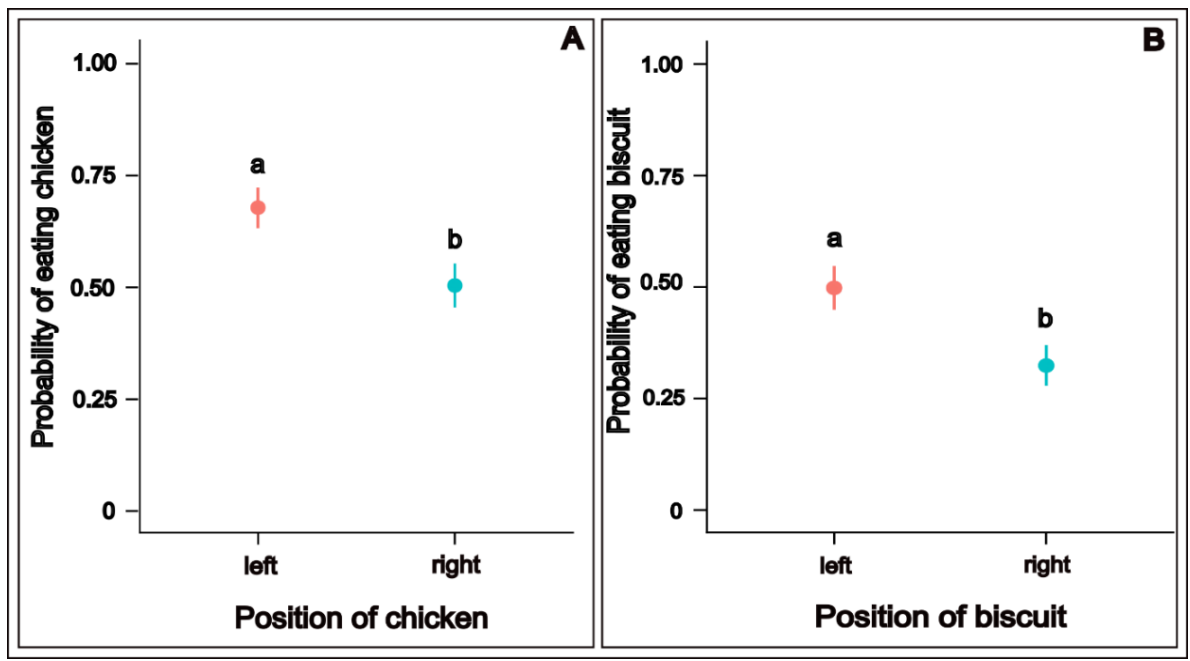

**Figure 5:** Model-estimated marginal means of the probability of eating food, A. chicken; B. biscuit, by a dog depending on the side they were kept at. The whiskers represent the 95% asymptotic confidence limits. The letters denote the presence of difference across the predictor, "positionfood".

### 3.7. How quickly do dogs respond to the set-up?

We analysed whether there is a difference in latency of approach in dogs between phases, age, and trials. We defined latency as the time taken from when the boxes were placed on the ground to the first sniff of one of the boxes by the dog. We ran a nested, mixed effects cox proportional hazards model with random effects of individual "dogids" nested within "bigplace". The maximum time provided to the dog to complete the experiment, 60s, was taken as the censored threshold value. The predictor variables were phase, age, and trial (t1, t2, and t3), and eatchoiceside(left/right). The "eatchoiceside" predictor denoted the side the dog approached and ate from. The censored variable was sniff which was binary (1: dog sniffed, 0: dog did not sniff)

(latency, sniff) ~ phase + age + trial1 + eatchoiceside + (1|bigplace/dogid)

The median latency across all conditions was 2s. The model estimated marginal means showed us that being a pup increases the "hazard" of sniffing [(adult – pup: -0.70, p-value < 0.05), (juvenile – pup: -0.52, p-value < 0.05)]. Additionally, the original phase is associated with a decreased likelihood of the set-up being sniffed as compared to others [(double black – original: 0.38, p-value < 0.05), (flipbox – original: 0.67, p-value < 0.05), (original – double transparent: -0.70, p-value < 0.05)]. This implies that pups are more likely to sniff earlier than adults and juveniles and dogs are expected to take longer to sniff in the original phase. Furthermore, dogs are likely to sniff earlier as the trials go on (t1 – t2: -0.16, p-value < 0.05), (t1 – t3: -0.30, p-value < 0.05), (t2 – t3: -0.14, p-value = 0.03). The latency is also lower when dogs approach the left side than when they do right (left – right: 0.131, 0-value < 0.05).

### 3.8. Does the presence (black) or absence (transparent) of colour influence the choice of a dog?

We tested a) whether the black or the transparent boxes themselves influenced the decision-making in dogs b) and if dogs of a certain age were attracted towards one box or the other. We ran an intercept-only model for the first case and a logistic regression for the second. Both models were hierarchical with random effects of "trial" nested within individual "dogid". Both the models had "firstsniffchoice" as its response variable (binary: black/transparent) that marked the choice of the dog. The binomial model had "age"- adult, juvenile, and, pup as predictor.

firstsniffchoice ~ 1 + (1 | dogid/trial)          . . . Intercept-only model

firstsniffchoice ~ 1 + age + (1 | dogid/trial)    . . . Binomial model

The results of the regression in both cases showed that either of the boxes had an equally likely chance to be chosen by the dogs (p-value > 0.05). Age had no effect on the choice either (p > 0.05).

## 4. Discussion

We investigated how dogs processed and prioritized visual and olfactory signals in the context of foraging. While pet dogs preferred vision-based strategies and olfaction has been suggested to be primary for free-ranging dogs when it comes to foraging, we found limited evidence of sensory prioritization or hierarchy between either when dogs were made to choose one over the other. We found that dogs initially sampled the boxes randomly across all phases, regardless of available cues, validating our hypothesis. This is in line with previous results and theories where animals (among them dogs) initiate sampling randomly in information constrained scenarios [59], [73]. Even across trials, dogs continue to sniff randomly, except for juveniles in the original and flip phases, who show a preference for the chicken box and biscuit box (both black), respectively.

Furthermore, dogs of all age groups show a preference for eating from the chicken box (black) in the original phase and adults in the flip phase also prioritise the chicken box (transparent). This hints at the prioritization of olfactory cue for finding their preferred food (chicken), which can be supplemented by visual signals. But the lack of similar results from the other two phases (selecting chicken in double black and double transparent phases) prevents us from stating this conclusively. Although limited, the dominance of olfaction in foraging tasks has been previously observed in free-ranging dogs but is in contradiction to pet dogs [48], [57], [58], [59]. The lack of a clear choice in the achromatic phase (double black) is also observed when dogs sniff both boxes. It is possible that the similar colours and shape of both the boxes in those phases impeded their decision-making. Animals have shown difficulty or inability to discriminate between shades of same achromatic signals [74], [75]. Frugivorous primates have difficulty detecting achromatic fruit signals because of large achromatic variance in foliage [76]. In dogs, it is known that they can discriminate brightness differences of spatially separated achromatic visual stimuli, a condition that may have been lacking in the double black phase [77]. In fact, colour cues, and specifically their differences are important visual signals that dogs use to discriminate between objects [78]. Furthermore, the illumination conditions might have also played a role in visual perception, and the transparent boxes might not have appeared as completely transparent to them, especially when perceived from a distance The absence of a choice in the double black and transparent phases would then highlight the importance of vision in foraging decisions.

Our results also suggest an alternate reason for the lack of a choice, optimal or otherwise in many instances. As observed, dogs moved in quickly towards the set-up with a median time of 2s, with this time getting further reduced in subsequent trials. Additionally, we also found that a majority of the dogs initiated eating without sampling both the boxes. This led to them missing out on their preferred food, but may not have been viewed so by the dog. While getting more and better information would have increased cue reliability and discrimination among the available options, the act of gathering information is also costly in terms of energy, time, loss of opportunities and exposure to risk [79]. In noisy and risky environments, like the ones occupied by free-ranging dogs, long term success and survival favours speed over accuracy. While this might lead to less-than-optimal results or incorrect information, the risk of potential dangers is also low. This speed-accuracy trade off strategy is especially favourable if the cost associated with the error is low. For example, bees are known to actively spend less effort and time at colour discrimination tasks when cost of error is low, like lower nectar rewards, but improve their response time and choice accuracy when errors are penalised [80], [81], [82]. In fact, it was found that bees that made more mistakes but made quicker decisions collected more nectar than their slow-but-accurate counterparts [83]. The markings of such similar speed-accuracy trade-off decisions can be seen in the free-ranging dogs too. From their point-of-view, the cost of making an error in their foraging decision-making is a less-preferred food. While, this might seem counter-intuitive to us, this should not be considered as a limitation of the sensory discrimination abilities of the dogs; this is rather an act of survival in a challenging environment.

Animals often devise strategies that get them the highest possible reward at the lowest-possible cost. The highest reward may not always be the best. Such "good-enough" strategy of intermediate selectivity was earlier highlighted in the fallow deer [84]. Similar behaviour was observed in mice too. Given the kind of heterogenous and unpredictable environment the free-ranging dog occupies with high risk and lost opportunity costs, the "good-enough" strategy seems apt for decision-making, especially since fresh, raw chicken pieces are rarely encountered in roadside garbage dumps and dustbins, common feeding spots of free-ranging dogs. However, we must not mistake this lack of discrimination as an actual inability to discriminate. Dogs that sniffed both boxes before eating chose to eat from the chicken-containing box and followed the Sniff-&-Snatch strategy to do so in the dichromatic phases [57], [85]. Thus, it maybe hypothesized that in the presence of clear, unambiguous signals dogs can quickly find their preferred food but in their absence, they settle for whatever they can get their paws on, displaying a labile response to available cues.

Another reason that may have hindered the decision-making in terms of food preference is the left-bias shown by the dogs. Although, the experimenters made sure that the boxes were randomly placed and the lack of unintentional side bias was also validated post-hoc, whichever food was placed on the left side had a higher likelihood of being eaten by the dog. This was in contradiction to previous studies in dogs looking at behavioural asymmetries as a measure of emotional states that linked the right hemisphere/left bias to alarming sensory stimuli and left hemisphere/right bias to positive, prosocial stimuli. Moreover, in a positive anticipatory situation where a dog could see the food before inspecting, they displayed a right eye/left hemisphere bias [51], [52], [53], [86], [87], [88]. A plausible explanation for the contradictory behaviour could come from looking at the visuospatial bias in humans and birds, rather than from the emotional state standpoint. In humans, this asymmetry derives from the right hemispheric dominance in controlling spatial attentional resources[89]. Hence, humans primarily attend to objects on the left side of space[90]. Birds also show a strong leftward bias in the spatial distribution of attention for food detection[27]. Additionally, the right hemisphere of the avian brain decodes relational spatial information. [2]. Keeping these findings in consideration, we may hypothesize that bereft of extreme emotional states, as seen in the prior studies, the visual encoding of spatial information has a leftward bias in dogs. Furthermore, the heterogenous and unpredictable nature of food availability, both in terms of space and time, make it more likely that a dog might make use of relational spatial information rather than specific visual landmarks and absolute metric distances, explaining their leftward choice. We also found that dogs that chose left were quicker to approach the set-up, adding further support that hemispheric compartmentalization and behavioural asymmetry lead to decision efficiency.

Pups were quicker to approach the set up and sniff one of the boxes than the adults and juveniles. Similar differential behaviour has been seen by pups in other studies too. This has been attributed to the fact that juveniles and adults are more cautious of humans because of their negative experience with them [91]. Moreover, young dogs go through a critical period of socialization when they readily accept human interaction, followed by a marked period of fear when new humans are avoided [92], [93]. This also implicates the role of learning and experience in the observed behavioural plasticity.  The dogs did not show preference on the basis of box colour. This assures us of the fact that the preference towards the black or transparent boxes in any of the phases had nothing to do with the box property itself and the approach and eventual eating was based on the available food cues.

While the study examined the interplay of sensory and internal factors against a backdrop of foraging strategies in ecologically relevant, multisensorial, natural environments, it presented its own difficulties. We were unable to present the set-up to the dogs in a controlled environment. The visual properties of the background and the surrounding noise of the environment might have had an effect on the study. We could not control for the direction of sunlight or how it interacted with the boxes nor the foliage or background against which the boxes were kept. All of that could have interfered with their visual information perception and subsequent processing. However, we need to keep in mind that the free-ranging dogs encounter such noisy backgrounds while scavenging in their daily lives. Furthermore, we could not completely separate the visual and olfactory modalities in our design (the transparent box), preventing us from examining the degree of effect of each signal in a dog's decision. While we used disposable gloves to prevent food signals cross contamination and cleaned the boxes between experiments, we could not control for any lingering chemical cues that may have been left on the boxes by previous dogs. Moreover, our study found a population-level bias towards left side approach when it comes to a choice task but we could not test the variation of its lateralization from one individual to another, which could also have affected outcome of the trials.

In conclusion, the current study provides some novel insights into the search and foraging strategy of free-ranging dogs. We discovered that dogs can use olfactory cue to search for their hidden, but preferred food. While the importance of visual sense was understood from what the dogs did not do in the absence of unambiguous cues, we were only able to establish its role in a limited manner and cannot clarify whether the two cues worked on a hierarchical or an integrated basis during foraging.

We still have much to understand about the interaction effects of real-world noise and dog senses in the foraging context. Our study also highlights the dynamic nature of a dog's foraging strategy wherein they find out the highest quality food when possible and settle for intermediate or low nutrient food, favouring speed of consumption over quality, when they cannot. Our study shows a strong leftward visuospatial population bias in the context of food detection. To our knowledge, this is the first study in free-ranging dogs showing that lateralization exists in the population and needs to be taken into account when designing choice tests. Further studies need to identify the inter-individual behavioural asymmetries in these dogs and control for those in study designs. Overall, in multi sensorial settings, visual and olfactory cues seem to be working in tandem with environmental signals and internal perceptual and cognitive elements. In the context of foraging, decisions and choices are complex phenomena, involve multiple modulators and may not be explained simply as a consequence of following one or the other cue.

**Author contributions**

RSa and ABha: project administration, and writing—reviewing and editing. RSa: conceptualization, formal analysis, methodology, visualization, and writing original draft. RSa, TsP, SM, AL: Data curation. RSa, TsP, SM, SN, AL, AkB, AGP, NW, NJ, SJ, SB, RA, TP, SMa: Investigation, ABha: funding acquisition, resources, supervision, and validation

**Ethics statement**

Ethical review and approval were not required for the animal study because the experiment did not involve any invasive procedure, and the food provided to the dogs were fit for human consumption. Dog feeding on streets is permitted by Prevention of Cruelty to Animals Act 1960 of the Parliament, and this experimental protocol did not need any additional clearance from the Institute ethics committee, as it did not violate the law.

**Conflict of interest**

The authors declare that the research was conducted in the absence of any commercial or financial relationships that could be construed as a potential conflict of interest.

**Funding**

RSa was supported by IISER Kolkata Institute fellowship. TsP was supported by SVMCM scholarship, Govt. of West Bengal and PhD fellowship of the University Grants

Commission (UGC) India. SM was supported by UGC PhD fellowship. SN was supported by the INSPIRE PhD fellowship, Department of Science and Technology, India. NW was funded by the KVPY fellowship, Govt. of India. This study was supported by IISER Kolkata ARF and the Janaki Ammal National Women Bioscientist Award (BT/HRD/NBA-NWB/39/2020-21 (YC-1)) of the Department of Biotechnology, India.

**References**

[1] D. Rubene, M. Leidefors, V. Ninkovic, S. Eggers, and M. Low, "Disentangling olfactory and visual information used by field foraging birds," *Ecol Evol*, vol. 9, no. 1, pp. 545–552, Jan. 2019, doi: 10.1002/ece3.4773.

[2] L. Tommasi and G. Vallortigara, "Encoding_of_geometric_and_landmark_information in the left and right hemisphere," *Behavioral Neuroscience*, vol. 115, pp. 602–613, 2001.

[3] C. C. Croney, K. M. Adams, C. G. Washington, and W. R. Stricklin, "A note on visual, olfactory and spatial cue use in foraging behavior of pigs: Indirectly assessing cognitive abilities," *Appl Anim Behav Sci*, vol. 83, no. 4, pp. 303–308, Oct. 2003, doi: 10.1016/S0168-1591(03)00128-X.

[4] T. Gilad, O. Bahar, M. Hasan, A. Bar, A. Subach, and I. Scharf, "The combined role of visual and olfactory cues in foraging by Cataglyphis ants in laboratory mazes," *Curr Zool*, vol. 69, no. 4, pp. 401–408, Aug. 2023, doi: 10.1093/cz/zoac058.

[5] M. Müller and R. Wehner, "Wind and sky as compass cues in desert ant navigation," *Naturwissenschaften*, vol. 94, no. 7, pp. 589–594, Jul. 2007, doi: 10.1007/s00114-007-0232-4.

[6] J. M. Plotnik, R. C. Shaw, D. L. Brubaker, L. N. Tiller, and N. S. Clayton, "Thinking with their trunks: Elephants use smell but not sound to locate food and exclude nonrewarding alternatives," *Anim Behav*, vol. 88, pp. 91–98, Feb. 2014, doi: 10.1016/j.anbehav.2013.11.011.

[7] V. Maselli, A. S. Al-Soudy, M. Buglione, M. Aria, G. Polese, and A. Di Cosmo, "Sensorial hierarchy in octopus vulgaris's food choice: Chemical vs. Visual," *Animals*, vol. 10, no. 3, Mar. 2020, doi: 10.3390/ani10030457.

[8] E. L. G. Legge, C. R. Madan, M. L. Spetch, and E. A. Ludvig, "Multiple cue use and integration in pigeons (Columba livia)," *Anim Cogn*, vol. 19, no. 3, pp. 581–591, May 2016, doi: 10.1007/s10071-016-0963-8.

[9] M. Senzaki *et al.*, "Sensory pollutants alter bird phenology and fitness across a continent," *Nature*, vol. 587, no. 7835, pp. 605–609, Nov. 2020, doi: 10.1038/s41586-020-2903-7.

[10] J. A. Shivik and L. Clark, "Carrion seeking in brown tree snakes: Importance of olfactory and visual cues," *Journal of Experimental Zoology*, vol. 279, no. 6, pp. 549–553, Dec. 1997, doi: 10.1002/(SICI)1097-010X(19971215)279:6<549::AID-JEZ2>3.0.CO;2-N.

[11] H. Raghuram, C. Thangadurai, N. Gopukumar, K. Nathar, and K. Sripathi, "The role of olfaction and vision in the foraging behaviour of an echolocating megachiropteran fruit bat, Rousettus leschenaulti (Pteropodidae)," *Mammalian Biology*, vol. 74, no. 1, pp. 9–14, Jan. 2009, doi: 10.1016/j.mambio.2008.02.008.

[12] R. A. Raguso and M. A. Willis, "Synergy between visual and olfactory cues in nectar feeding by naïve hawkmoths, Manduca sexta," *Anim Behav*, vol. 64, no. 5, pp. 685–695, 2002, doi: 10.1006/anbe.2002.4010.

[13] S. Potier, O. Duriez, A. Célérier, J. L. Liegeois, and F. Bonadonna, "Sight or smell: which senses do scavenging raptors use to find food?," *Anim Cogn*, vol. 22, no. 1, pp. 49–59, Jan. 2019, doi: 10.1007/s10071-018-1220-0.

[14] J. A. Teichroeb and C. A. Chapman, "Sensory information and associative cues used in food detection by wild vervet monkeys," *Anim Cogn*, vol. 17, no. 3, pp. 517–528, 2014, doi: 10.1007/s10071-013-0683-2.

[15] H. Jachmann, "Food Selection by Elephants in the 'Miombo' Biome, in Rela on to Leaf Chemistry," *Biochem Syst Ecol*, vol. 17, pp. 15–24, 1989.

[16] R. S. Stutz, B. M. Croak, N. Proschogo, P. B. Banks, and C. McArthur, "Olfactory and visual plant cues as drivers of selective herbivory," *Oikos*, vol. 126, no. 2, Feb. 2017, doi: 10.1111/oik.03422.

[17] G. A. Nevitt, M. Losekoot, and H. Weimerskirch, "Evidence for olfactory search in wandering albatross, Diomedea exulans," *Proceedings of the National Academy of Sciences*, vol. 105, no. 12, pp. 4576–4581, Mar. 2008, doi: 10.1073pnas.0709047105.

[18] G. Gottsberger and I. Silberbauer-Gottsberger, "Olfactory and Visual Attraction of Erioscelis emarginata (Cyclocephalini, Dynastinae) to the Inflorescences of Philodendron selloum (Araceae)," *Biotropica*, vol. 23, no. 1, pp. 23–28, 1991, Accessed: Apr. 25, 2025. [Online]. Available: http://www.jstor.org/stable/2388684?origin=JSTOR-pdf

[19] I. T. Baldwin, R. Halitschke, A. Paschold, C. C. Von Dahl, and C. A. Preston, "Volatile signaling in plant-plant interactions: 'Talking trees' in the genomics era," Feb. 10, 2006. doi: 10.1126/science.1118446.

[20] H. Warburton and G. Mason, "Is out of sight out of mind? The effects of resource cues on motivation in mink, Mustela vison," *Anim Behav*, vol. 65, no. 4, pp. 755–762, Apr. 2003, doi: 10.1006/anbe.2003.2097.

[21] J. P. Croxall and P. A. Prince, "Dead or alive, night or day: How do albatrosses catch squid?," *Antarct Sci*, vol. 6, no. 2, pp. 155–162, 1994, doi: 10.1017/S0954102094000246.

[22] D. Rubene, M. Low, and A. Brodin, "Birds differentially prioritize visual and olfactory foraging cues depending on habitat of origin and sex," *R Soc Open Sci*, vol. 10, no. 2, Feb. 2023, doi: 10.1098/rsos.221336.

[23] S. Y. Yang, B. A. Walther, and G. J. Weng, "Stop and smell the pollen: The role of olfaction and vision of the oriental honey buzzard in identifying food," *PLoS One*, vol. 10, no. 7, Jul. 2015, doi: 10.1371/journal.pone.0130191.

[24] B. Wist, C. Stolter, and K. H. Dausmann, "Sugar addicted in the city: impact of urbanisation on food choice and diet composition of the Eurasian red squirrel (Sciurus vulgaris)," *Journal of Urban Ecology*, vol. 8, no. 1, 2022, doi: 10.1093/jue/juac012.

[25] M. Murray, A. Cembrowski, A. D. M. Latham, V. M. Lukasik, S. Pruss, and C. C. St Clair, "Greater consumption of protein-poor anthropogenic food by urban relative to rural coyotes increases diet breadth and potential for human-wildlife conflict," *Ecography*, vol. 38, no. 12, pp. 1235–1242, Dec. 2015, doi: 10.1111/ecog.01128.

[26] K. Karenina, A. Giljov, T. Ivkovich, and Y. Malashichev, "Evidence for the perceptual origin of right-sided feeding biases in cetaceans," *Anim Cogn*, vol. 19, no. 1, pp. 239–243, Jan. 2016, doi: 10.1007/s10071-015-0899-4.

[27] B. Diekamp, L. Regolin, O. Güntürkün, and G. Vallortigara, "A left-sided visuospatial bias in birds," *Current Biology*, vol. 15, pp. R372–R373, 2005, [Online]. Available: http://www.current-

[28] L. J. Rogers, P. Zucca, and G. Vallortigara, "Advantages of having a lateralized brain," *Proceedings of the Royal Society B: Biological Sciences*, vol. 271, no. SUPPL. 6, Dec. 2004, doi: 10.1098/rsbl.2004.0200.

[29] O. Güntürkün, B. Diekamp, M. Manns, F. Nottelmann, A. Schwarz, and M. Skiba, "Asymmetry pays: visual lateralization improves discrimination success in pigeons," 2000.

[30] L. J. Rogers, P. Zucca, and G. Vallortigara, "Advantages of having a lateralized brain," *Proceedings of the Royal Society B: Biological Sciences*, vol. 271, no. SUPPL. 6, Dec. 2004, doi: 10.1098/rsbl.2004.0200.

[31] M. Vinassa *et al.*, "Palatability assessment in horses in relation to lateralization and temperament," *Appl Anim Behav Sci*, vol. 232, Nov. 2020, doi: 10.1016/j.applanim.2020.105110.

[32] Z. Polgár, M. Kinnunen, D. Újváry, Á. Miklósi, and M. Gácsi, "A test of canine olfactory capacity: Comparing various dog breeds and wolves in a natural detection task," *PLoS One*, vol. 11, no. 5, May 2016, doi: 10.1371/journal.pone.0154087.

[33] G. A. A. Schoon ', "Scent identification lineups by dogs ( Cunis familiaris) : experimental design and forensic application," 1996.

[34] D. G. Moulton, "Minimum odorant concentrations detectable by the dog and their implications for olfactory receptor sensitivity," in *Chemical signals in vertebrates*, Boston: Springer, 1977, pp. 455–464.

[35] J. Bräuer and B. Vidal Orga, "Why wolves became dogs: Interdisciplinary questions on domestication," in *Dogs, Past and Present*, I. Fiore and F. Lugli, Eds., Archaeopress Archaeology, 2023.

[36] V. Ruusila and M. Pesonen, "Interspecific cooperation in human (Homo sapiens) hunting: the benefits of a barking dog," *Ann Zool Fennici*, vol. 41, no. 4, pp. 545–549, 2004.

[37] I. Gazit and J. Terkel, "Domination of olfaction over vision in explosives detection by dogs," *Appl Anim Behav Sci*, vol. 82, no. 1, pp. 65–73, Jun. 2003, doi: 10.1016/S0168-1591(03)00051-0.

[38] T. Jezierski *et al.*, "Efficacy of drug detection by fully-trained police dogs varies by breed, training level, type of drug and search environment," *Forensic Sci Int*, vol. 237, pp. 112–118, 2014, doi: 10.1016/j.forsciint.2014.01.013.

[39] L. R. Bijland, M. K. Bomers, and Y. M. Smulders, "Smelling the Diagnosis: A review on the use of scent in diagnosing disease," *The Journal of Medicine*, vol. 71, no. 6, 2013.

[40] A. Fick, "V. On liquid diffusion," *The London, Edinburgh, and Dublin Philosophical Magazine and Journal of Science*, vol. 10, no. 63, pp. 30–39, 1855.

[41] Z. Polgár, Á. Miklósi, and M. Gácsi, "Strategies used by pet dogs for solving olfaction-based problems at various distances," *PLoS One*, vol. 10, no. 7, Jul. 2015, doi: 10.1371/journal.pone.0131610.

[42] S. E. Byosiere, P. A. Chouinard, T. J. Howell, and P. C. Bennett, "What do dogs (Canis familiaris) see? A review of vision in dogs and implications for cognition research," Oct. 01, 2018, *Springer Science and Business Media, LLC*. doi: 10.3758/s13423-017-1404-7.

[43] P. E. Miller and C. J. Murphy, "Vision in Dogs," *J Am Vet Med Assoc*, vol. 207, no. 12, Dec. 1995.

[44] A. A. Kasparson, J. Badridze, and V. V. Maximov, "Colour cues proved to be more informative for dogs than brightness," *Proceedings of the Royal Society B: Biological Sciences*, vol. 280, no. 1766, Aug. 2013, doi: 10.1098/rspb.2013.1356.

[45] J. M. Baker, J. Morath, K. S. Rodzon, and K. E. Jordan, "A shared system of representation governing quantity discrimination in canids," *Front Psychol*, vol. 3, no. OCT, 2012, doi: 10.3389/fpsyg.2012.00387.

[46] M. E. M. Petrazzini and C. D. L. Wynne, "What counts for dogs (Canis lupus familiaris) in a quantity discrimination task?," *Behavioural Processes*, vol. 122, pp. 90–97, 2016.

[47] M. C. Wells and P. N. Lehner, "The relative importance of the distance senses in coyote predatory behaviour," 1978.

[48] J. Bräuer, V. Mann, and J. Erlacher, "Eyes or Nose: Domestic Dogs (Canis familiaris) Prefer Vision Over Olfaction When Searching for Food," *J Comp Psychol*, 2025, doi: 10.1037/com0000415.

[49] K. Guo, K. Meints, C. Hall, S. Hall, and D. Mills, "Left gaze bias in humans, rhesus monkeys and domestic dogs," *Anim Cogn*, vol. 12, no. 3, pp. 409–418, May 2009, doi: 10.1007/s10071-008-0199-3.

[50] M. Siniscalchi, R. Sasso, A. M. Pepe, G. Vallortigara, and A. Quaranta, "Dogs turn left to emotional stimuli," *Behavioural Brain Research*, vol. 208, no. 2, pp. 516–521, Apr. 2010, doi: 10.1016/j.bbr.2009.12.042.

[51] M. Siniscalchi, R. Sasso, A. M. Pepe, S. Dimatteo, G. Vallortigara, and A. Quaranta, "Sniffing with the right nostril: Lateralization of response to odour stimuli by dogs," *Anim Behav*, vol. 82, no. 2, pp. 399–404, Aug. 2011, doi: 10.1016/j.anbehav.2011.05.020.

[52] A. Quaranta, M. Siniscalchi, and G. Vallortigara, "Asymmetric tail-wagging responses by dogs to different emotive stimuli," Mar. 20, 2007. doi: 10.1016/j.cub.2007.02.008.

[53] T. Simon, A. Wilkinson, E. Frasnelli, K. Guo, and D. S. Mills, "Lateralized behaviour in dogs during positive anticipation," *Anim Behav*, vol. 216, pp. 155–173, Oct. 2024, doi: 10.1016/j.anbehav.2024.08.005.

[54] J. Serpell, "From paragon to pariah: Some reflections on human attitudes to dogs.," in *The domestic dog : its evolution, behaviour, and interactions with people*, J. Serpell, Ed., • Cambridge ; New York: Cambridge University Press, 1995, pp. 245–256.

[55] Thapar, *A history of India*, vol. 1. 1966. doi: 10.1017/CBO9781107415324.004.

[56] B. Debroy, *Sarama and Her Children_ The Dog in Indian Myth*. Penguin Books India, 2008.

[57] A. Bhadra, D. Bhattacharjee, M. Paul, and A. Bhadra, "The Meat of the Matter: A thumb rule for scavenging dogs?," *Ethol Ecol Evol*, vol. 9370, no. September, 2015, doi: 10.1080/03949370.2015.1076526.

[58] R. Sarkar, S. Sau, and A. Bhadra, "Scavengers can be choosers: A study on food preference in free-ranging dogs," *Appl Anim Behav Sci*, vol. 216, no. April, pp. 38–44, 2019, doi: 10.1016/j.applanim.2019.04.012.

[59] R. Sarkar *et al.*, "Eating smart: Free-ranging dogs follow an optimal foraging strategy while scavenging in groups," *Front Ecol Evol*, vol. 11, 2023, doi: 10.3389/fevo.2023.1099543.

[60] S. Biswas *et al.*, "Scavengers in the human-dominated landscape: An experimental study," *Philosophical Transactions of the Royal Society B: Biological Sciences*, vol. 379, no. 1909, Jul. 2024, doi: 10.1098/rstb.2023.0179.

[61] A. Banerjee and A. Bhadra, "The more the merrier: Dogs can assess quantities in food-choice tasks," *Curr Sci*, vol. 117, no. 6, pp. 1095–1100, 2019, doi: 10.18520/cs/v117/i6/1095-1100.

[62] D. Bhattacharjee, S. Sau, J. Das, and A. Bhadra, "Does novelty influence the foraging decisions of a scavenger?," *PeerJ*, vol. 12, 2024, doi: 10.7717/peerj.17121.

[63] A. Roy *et al.*, "Ready, set, yellow! color preference of Indian free-ranging dogs," *Anim Cogn*, vol. 28, no. 1, Dec. 2025, doi: 10.1007/s10071-024-01928-9.

[64] S. Sen Majumder *et al.*, "To be or not to be social: Foraging associations of free-ranging dogs in an urban ecosystem," *Acta Ethol*, vol. 17, no. 1, pp. 1–8, 2014, doi: 10.1007/s10211-013-0158-0.

[65] R Core Team, "R: A Language and Environment for Statistical Computing," 2023, *R Foundation for Statistical Computing, Vienna*: 4.3.1.

[66] D. Bates, M. Mächler, B. M. Bolker, and S. C. Walker, "Fitting linear mixed-effects models using lme4," *J Stat Softw*, vol. 67, no. 1, Oct. 2015, doi: 10.18637/jss.v067.i01.

[67] M. Elff, "Package 'mclogit,'" 2022. Accessed: Apr. 16, 2025. [Online]. Available: http://mclogit.elff.eu,https://github.com/melff/mclogit/

[68] Russell V. Lenth, "emmeans: Estimated Marginal Means, aka Least-Squares Means," 2024, R package version 1.10.0. Accessed: Mar. 10, 2024. [Online]. Available: https://CRAN.R-project.org/package=emmeans

[69] S. R. Searle, F. M. Speed, and G. A. Milliken, "Population marginal means in the linear model: An alternative to least squares means," *American Statistician*, vol. 34, no. 4, pp. 216–221, 1980, doi: 10.1080/00031305.1980.10483031.

[70] M. Jafari and N. Ansari-Pour, "Why, when and how to adjust your P values?," *Cell J*, vol. 20, no. 4, pp. 604–607, 2019, doi: 10.22074/cellj.2019.5992.

[71] S. Nakagawa, "A farewell to Bonferroni: The problems of low statistical power and publication bias," Nov. 2004. doi: 10.1093/beheco/arh107.

[72] Therneau T, "A Package for Survival Analysis in R," 2023, R package version 3.5-5. Accessed: Apr. 28, 2024. [Online]. Available: https://CRAN.R-project.org/package=survival

[73] F. Bartumeus, D. Campos, W. S. Ryu, R. Lloret-Cabot, V. Méndez, and J. Catalan, "Foraging success under uncertainty: search tradeoffs and optimal space use," *Ecol Lett*, vol. 19, no. 11, pp. 1299–1313, Nov. 2016, doi: 10.1111/ele.12660.

[74] A. Kelber, M. Vorobyev, and D. Osorio, "Animal colour vision - Behavioural tests and physiological concepts," Feb. 2003. doi: 10.1017/S1464793102005985.

[75] H. M. Schaefer, D. J. Levey, V. Schaefer, and M. L. Avery, "The role of chromatic and achromatic signals for fruit detection by birds," *Behavioral Ecology*, vol. 17, no. 5, pp. 784–789, Sep. 2006, doi: 10.1093/beheco/arl011.

[76] P. Sumner and J. D. Mollon, "Chromaticity as a signal of ripeness in fruits taken by primates," *Journal of Experimental Biology*, vol. 203, no. 13, pp. 1987–2000, Jun. 2000.

[77] G. Pretterer, H. Bubna-Littitz, G. Windischbauer, C. Gabler, and U. Griebel, "Brightness discrimination in the dog," *J Vis*, vol. 4, no. 3, pp. 241–249, Apr. 2004, doi: 10.1167/4.3.10.

[78] A. A. Kasparson, J. Badridze, and V. V. Maximov, "Colour cues proved to be more informative for dogs than brightness," *Proceedings of the Royal Society B: Biological Sciences*, vol. 280, no. 1766, Aug. 2013, doi: 10.1098/rspb.2013.1356.

[79]   A. Sih, "Prey uncertainty and the balancing of antipredator and feeding needs," *Am Nat*, vol. 139, no. 5, pp. 1052–1069, May 1992, [Online]. Available: http://www.journals.uchicago.edu/t-and-c

[80]   L. Chittka, A. G. Dyer, F. Bock, and A. Dornhaus, "Bees trade off foraging speed for accuracy," *Nature*, vol. 424, pp. 388–388, Jul. 2003.

[81]   L. Chittka, P. Skorupski, and N. E. Raine, "Speed-accuracy tradeoffs in animal decision making," Jul. 2009. doi: 10.1016/j.tree.2009.02.010.

[82]   A. G. Dyer and L. Chittka, "Bumblebees (Bombus terrestris) sacrifice foraging speed to solve difficult colour discrimination tasks," *J Comp Physiol A Neuroethol Sens Neural Behav Physiol*, vol. 190, no. 9, pp. 759–763, Sep. 2004, doi: 10.1007/s00359-004-0547-y.

[83]   J. G. Burns, "Impulsive bees forage better: The advantage of quick, sometimes inaccurate foraging decisions," *Anim Behav*, vol. 70, no. 6, Dec. 2005, doi: 10.1016/j.anbehav.2005.06.002.

[84]   U. Alm, B. Birgersson, and O. Leimar, "The effect of food quality and relative abundance on food choice in fallow deer," *Anim Behav*, vol. 64, no. 3, pp. 439–445, 2002, doi: 10.1006/anbe.2002.3057.

[85]   A. Bhadra and A. Bhadra, "Preference for meat is not innate in dogs," *J Ethol*, vol. 32, no. 1, pp. 15–22, 2014, doi: 10.1007/s10164-013-0388-7.

[86]   M. Siniscalchi, S. D'Ingeo, S. Fornelli, and A. Quaranta, "Lateralized behavior and cardiac activity of dogs in response to human emotional vocalizations," *Sci Rep*, vol. 8, no. 1, Dec. 2018, doi: 10.1038/s41598-017-18417-4.

[87]   M. Siniscalchi, S. d'Ingeo, and A. Quaranta, "Orienting asymmetries and physiological reactivity in dogs' response to human emotional faces," *Learn Behav*, vol. 46, no. 4, pp. 574–585, Dec. 2018, doi: 10.3758/s13420-018-0325-2.

[88]   A. Racca, K. Guo, K. Meints, and D. S. Mills, "Reading faces: Differential lateral gaze bias in processing canine and human facial expressions in dogs and 4-year-old children," *PLoS One*, vol. 7, no. 4, Apr. 2012, doi: 10.1371/journal.pone.0036076.

[89] A. C. Nobre, J. T. Coull, P. Maquet, C. D. Frith, R. Vandenberghe, and M. M. Mesulam, "Orienting Attention to Locations in Perceptual Versus Mental Representations," *J Cogn Neurosci*, vol. 16, no. 3, pp. 363–373, 2004.

[90] B. Uttl and C. Pilkenton-Taylor, "Letter cancellation performance across the adult life span," *Clinical Neuropsychologist*, vol. 15, no. 4, pp. 521–530, 2001, doi: 10.1076/clin.15.4.521.1881.

[91] D. Bhattacharjee *et al.*, "Free-ranging dogs show age related plasticity in their ability to follow human pointing," *PLoS One*, vol. 12, no. 7, pp. 1–17, 2017, doi: 10.1371/journal.pone.0180643.

[92] K. Lord, "A Comparison of the Sensory Development of Wolves (Canis lupus lupus) and Dogs (Canis lupus familiaris)," *Ethology*, vol. 119, no. 2, pp. 110–120, Feb. 2013, doi: 10.1111/eth.12044.

[93] D. G. Freedman, J. A. King, and E. Orville, "Critical period in the social development of dogs," *Science (1979)*, vol. 133, no. 3457, pp. 1016–1017, 1961.